\documentclass[epj]{svjour}
\usepackage{graphics}
\usepackage{tensor}
\usepackage{amssymb}
\usepackage{feynmp}
\usepackage{psfrag}

\def\Im{\mathop{\rm Im}\nolimits}

\begin{document}
\begin{fmffile}{diagrams}

\title{Transport coefficients in Chiral Perturbation Theory}

\author{D. Fern\'andez-Fraile\thanks{Electronic address: danfer@fis.ucm.es}
\and A. G\'omez Nicola\thanks{Electronic address: gomez@fis.ucm.es}
}

\institute{Departamentos de  F\'\i sica Te\'orica I y II, Universidad Complutense, 28040 Madrid, Spain.}

\date{Oct. 15, 2006}

\abstract{We present recent results on the calculation of transport
coefficients for a pion gas at zero chemical potential in Chiral
Perturbation Theory  using Linear Response Theory. More precisely,
we show the behavior of DC conductivity and shear viscosity at low
temperatures. To compute transport coefficients, the standard power
counting of ChPT has to be modified. The effects derived from
imposing unitarity are also analyzed. As  physical applications in
Relativistic Heavy Ion Collisions, we show the relation of the DC
conductivity to soft-photon production and  phenomenological effects
related to a nonzero shear viscosity. In addition, our values for
the shear viscosity to entropy ratio satisfy the KSS bound.
}

\maketitle

\section{Introduction}
We are interested in studying transport properties in a pion gas
with zero baryon density. Pions are the lightest hadronic degrees of
freedom and their dynamics is well described for low-enough energies
and temperatures by Chiral Perturbation Theory (ChPT) \cite{Gasser}.
ChPT is a low-energy expansion performed in terms of $p^2$ ($p$
represents a momentum, a mass or a temperature) against some scale
$\Lambda_\chi^2\sim (1\ \mathrm{GeV})^2$. However, as we shall
see, the computation of transport coefficients in ChPT is
intrinsically non-perturbative, leading to a modification of the
standard ChPT power counting scheme to take into account collisions
in the plasma (i.e. a non-zero pion width).

Considering small external perturbations which put the system
slightly out of equilibrium, we can employ Linear Response Theory
(LRT) \cite{LeBellac,Jeon,Valle}, in which a transport coefficient
$\mathfrak{t}$ is given by:
\begin{equation}\label{coeff}
{\mathfrak{t}}= C \lim_{q^0\rightarrow 0^+}\lim_{|\vec{q}|\rightarrow 0^+}\frac{\partial\rho(q^0,\vec{q})}{\partial q^0}\ ,
\end{equation}
with $C$ a constant and $\rho$ the spectral density of a current-current correlator.

\begin{figure}[!ht]
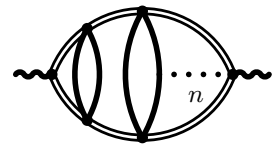

\begin{center}
\resizebox{0.20\textwidth}{!}{
\begin{fmfchar*}(90,40)
\fmfpen{thick}
\fmf{phantom,left=.7,tension=.1}{vw,ve,vw}
\fmfleft{w}
\fmfright{e}
\fmf{boson}{w,vw}
\fmf{boson}{ve,e}
\fmffreeze
\fmfipath{pn,ps,camino}\fmfipair{vn,vs}\fmfipair{yn,zs}
\fmfiequ{pn}{vpath (__vw, __ve)}
\fmfiequ{ps}{vpath (__ve, __vw)}
\fmfiequ{vn}{point .5*length(pn) of pn}
\fmfiequ{vs}{point .5*length(ps) of ps}
\fmfiequ{yn}{point .25*length(pn) of pn}
\fmfiequ{zs}{point .75*length(ps) of ps}
\fmfi{double,width=thin}{subpath (0,.5)*length(pn) of pn}
\fmfi{double,width=thin}{subpath (.5,1)*length(pn) of pn}
\fmfi{double,width=thin}{subpath (0,.5)*length(pn) of ps}
\fmfi{double,width=thin}{subpath (.5,1)*length(pn) of ps}
\fmfi{plain}{vn{dir -60}..vs}
\fmfi{plain}{vs{dir 120}..vn}
\fmfi{plain}{yn{dir -60}..zs}
\fmfi{plain}{zs{dir 120}..yn}
\fmfipair{puntoin,puntofin}
\fmfiequ{camino}{vn{dir -60}..vs}
\fmfiequ{puntoin}{point .5*length(camino) of camino}
\fmfiequ{puntofin}{point length(pn) of pn}
\fmfi{dots,label=$n$}{puntoin--puntofin}
\fmfv{decor.shape=circle,decor.filled=full,decor.size=1thick}{ve}
\fmfv{decor.shape=circle,decor.filled=full,decor.size=1thick}{vw}
\fmfiv{decor.shape=circle,decor.filled=full,decor.size=1thick}{vn}
\fmfiv{decor.shape=circle,decor.filled=full,decor.size=1thick}{vs}
\fmfiv{decor.shape=circle,decor.filled=full,decor.size=1thick}{yn}
\fmfiv{decor.shape=circle,decor.filled=full,decor.size=1thick}{zs}
\end{fmfchar*}
}
\end{center}
\caption{Generic ladder diagram with $n$ rungs ($n\geq 1$). Double lines represent particles with $\mathnormal{\Gamma}\neq 0$.}\label{laddiag}
\end{figure}

\begin{figure}[!ht]
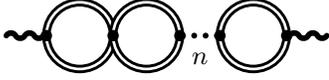

\begin{center}
\resizebox{0.25\textwidth}{!}{
\begin{fmfchar*}(110,40)
\fmfpen{thick}
\fmfleft{i}
\fmfright{o}
\fmf{photon}{i,v1}
\fmf{photon}{v5,o}
\fmf{double,width=thin,left,tension=.3}{v1,v2,v1}
\fmf{double,width=thin,left,tension=.3}{v2,v3,v2}
\fmf{dots,label=$n$}{v3,v4}
\fmf{double,width=thin,left,tension=.3}{v4,v5,v4}
\fmfv{decor.shape=circle,decor.filled=full,decor.size=1thick}{v1}
\fmfv{decor.shape=circle,decor.filled=full,decor.size=1thick}{v5}
\fmfv{decor.shape=circle,decor.filled=full,decor.size=1thick}{v2}
\fmfv{decor.shape=circle,decor.filled=full,decor.size=1thick}{v3}
\fmfv{decor.shape=circle,decor.filled=full,decor.size=1thick}{v4}
\end{fmfchar*}
}
\end{center}
\vspace{-0.2cm}\caption{Generic bubble diagram with $n$ bubbles
($n\geq 1$). Double lines represent particles with
$\mathnormal{\Gamma}\neq 0$.}\label{bubdiag}
\end{figure}

In the calculation of transport coefficients, there is a dominant
contribution coming from products of the kind
$G_\mathrm{A}G_\mathrm{R}\propto 1/\Gamma$ ($G_\mathrm{A}$ and
$G_\mathrm{R}$ are the advanced and retarded propagators
respectively), called \emph{pinching poles} \cite{Jeon,Valle}, where
$\mathnormal{\Gamma}$ is the particle width (which is perturbatively
small: $\mathnormal{\Gamma}=\mathcal{O}(p^5)$ in ChPT \cite{Goity}).
From Kinetic Theory (KT), a behavior $\sim 1/\mathnormal{\Gamma}$
for transport coefficients is also expected, since
$1/\mathnormal{\Gamma}$ represents the mean time between two
collisions of the particles in the plasma. Each of these pinching
poles comes from a pair of lines in a Feynman diagram contributing
to the spectral density that share the same four-momentum when the
limit in (\ref{coeff}) is taken. Accordingly, there are some classes
of diagrams potentially non-perturbative, being the most harmful the
type called \emph{ladder diagrams} (see Fig. \ref{laddiag}). Another
type, the so called \emph{bubble diagrams} (see Fig. \ref{bubdiag}),
which in principle would be the most harmful, for $n\geq 2$ turn out to give a
negligible contribution to the electrical conductivity and the shear
viscosity. Thus, the standard power counting scheme from ChPT has to
be modified to compute transport coefficients \cite{Conductivity}.
In the case of a ladder diagram with $n$ rungs, its contribution is
in principle of order $\mathcal{O}(p^{2n}Y^{n+1})$, where $Y$ is the
contribution from a single-bubble diagram. For $T\ll M_\pi\simeq 140\ \mathrm{MeV}$, a detailed analysis of the spectral function
\cite{Conductivity} shows that it contributes
$\mathcal{O}(p^{2n}Y)$. For $T\gtrsim M_\pi$, $Y$ becomes of order 1
or larger (due to unitarity effects) so that diagrams with one or
more rungs could be non-negligible in this regime. Also note that,
for high enough temperatures where typical momenta $p\sim T$,
diagrams with derivative vertices become potentially dominant
against those with constant vertices.

\section{Electrical conductivity}
\begin{figure}[ht]
\begin{center}
\psfrag{ylabel}[][][2]{$\displaystyle\frac{\sigma^{(0)}}{M_\pi}$}
\psfrag{xlabel}[][][2]{$T\ (\mathrm{MeV})$}
\psfrag{unit}[l][l][2]{unitarized $\mathcal{O}(p^4)$}
\psfrag{unitT}[l][l][2]{$T$-unitarized $\mathcal{O}(p^4)$}
\psfrag{p4}[l][l][2]{$\mathcal{O}(p^4)$}
\resizebox{0.38\textwidth}{!}{
  \includegraphics{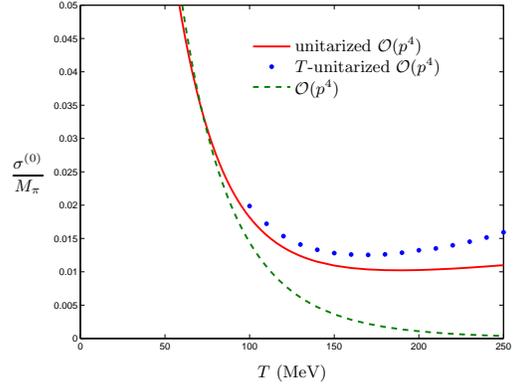}
}
\end{center}
\vspace{-0.3cm}\caption{Contribution to the electrical conductivity
from the single-bubble diagram as a function of  temperature. Three
curves for the pion width are shown: non-unitarized ${\cal O}(p^4)$
 (dashed), $T=0$ unitarized and finite $T$
unitarized.}\label{condunit}
\end{figure}

We are interested in the DC electrical conductivity, which
represents the response of the gas under the action of a constant
electric field. The expression for this transport coefficient in LRT
is:
\[
\sigma=-\lim_{q^0\rightarrow 0^+}\lim_{|\vec{q}|\rightarrow 0^+}\frac{\rho_\sigma(q^0,\vec{q})}{6q^0}\ ,
\]
with
\[
\rho_\sigma(q^0,\vec{q})=2\Im\mathrm{i}\int\mathrm{d}^4x\
\mathrm{e}^{\mathrm{i}q\cdot x}
\theta(t)\langle[{J}_i(x),{J}^i(0)]\rangle\ ,
\]
and $J^i$ is the electric current density in the gas. For this
transport coefficient, at $T\ll M_\pi$, $Y\simeq\sqrt{M_\pi/T}$. So
the contribution from a single-bubble diagram to the conductivity at
these temperature is $\sigma^{(0)}\simeq e^2M_\pi\sqrt{M_\pi/T}$,
where $e$ is the charge of the electron. The pion width is
calculated in ChPT \cite{Goity} from the pion-pion scattering
amplitudes. To consider the unitarity effects on the transport
coefficients, we impose unitarity on these amplitudes by means of
the Inverse Amplitude Method (IAM) \cite{iam}. This unitarized
scattering amplitudes have also been calculated at  finite
temperature \cite{thermalrho}. We have implemented unitarization in
the \emph{dilute gas approximation}, where we neglect ${\cal
O}(n_B^2)$ terms, with $n_B$ the Bose-Einstein distribution
evaluated at a pion energy. This is approximately valid for $T\leq
M_\pi$. In Fig. \ref{condunit} we represent the behavior with
temperature of the $\sigma^{(0)}$ contribution to conductivity. We
see that unitarity makes conductivity grow near $T=170\
\mathrm{MeV}$, as is expected in the QGP phase \cite{Arnold}. The
contribution from ladder diagrams with derivative vertices has to be
taken into account for temperatures $T\gtrsim M_\pi$.

%\begin{figure}[ht]
%\begin{center}
%\psfrag{ylabel}[][][2]{$\displaystyle\frac{\sigma^{(1)}}{\sigma^{(0)}}$}
%\psfrag{xlabel}[][][2]{$T\ (\mathrm{MeV})$}
%\psfrag{derunit}[l][l][2]{Derivative and unitarized (DGA)}
%\psfrag{derdg}[l][l][2]{Derivative (DGA)}
%\psfrag{constdg}[l][l][2]{Constant (DGA)}
%\resizebox{0.38\textwidth}{!}{
%  \includegraphics{plot2_2.eps}
%}
%\end{center}
%\vspace{-0.3cm}\caption{Comparison of the contributions from constant
%and derivative vertices in the dilute gas approximation (DGA)
%with and without unitarizing.}\label{derivplot}
%\end{figure}

\begin{figure}[!ht]
\begin{center}
\resizebox{0.38\textwidth}{!}{
  \includegraphics{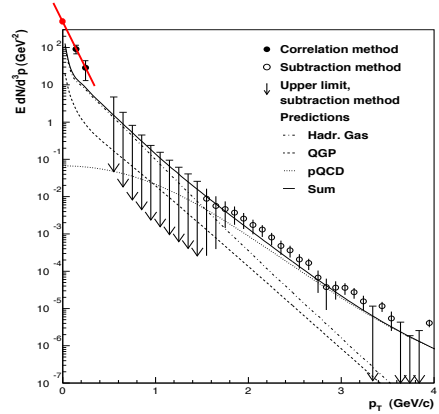}
}
\end{center}
\vspace{-0.3cm}\caption{Photon spectrum obtained by the experiment
WA98 \cite{WA98}. We show the linear extrapolation from the two lowest energy
points, which reaches a point at the origin compatible with our
estimate. Note that theoretical predictions tend to underestimate
the lowest energy points \cite{Rapp}.}\label{WA98plot}
\end{figure}

The DC conductivity is related to the zero-energy (soft) photon
spectrum emitted by the pion gas. Considering a pion gas produced
after a Relativistic Heavy Ion Collision which expands cylindrically
(Bjorken's hydrodynamical model), we obtain \cite{Conductivity} the
estimate $E\mathrm{d}N_\gamma/\mathrm{d}^3\vec{p}(p_T\rightarrow
0^+)\simeq 5.6\times 10^2\ \mathrm{GeV}^{-2}$ for the photon
spectrum at zero transverse momentum. In Fig. \ref{WA98plot} we see
that this estimate is compatible with the experimental results
obtained by the experiment WA98 \cite{WA98} and other theoretical
analysis \cite{Rapp}. This is reasonable since it is expected that
the hadronic contribution dominates at low $p_T$.

\section{Shear viscosity}
The expression for shear viscosity in LRT is
\[
\eta=\lim_{q^0\rightarrow 0^+}\lim_{|\vec{q}|\rightarrow 0^+}\frac{\rho_\eta(q^0,\vec{q})}{20q^0}\ ,
\]
where
\[
\rho_\eta(q^0,\vec{q})=2\Im\mathrm{i}\int\mathrm{d}^4x\
\mathrm{e}^{\mathrm{i}q\cdot
x}\theta(t)\langle[{\pi}_{ij}(x),{\pi}^{ij}(0)]\rangle\ ,
\]
and $\pi_{ij}$ is the traceless part of the spatial energy-momentum
tensor, i.e.
\[
\pi_{ij}\equiv T_{ij}-\frac{1}{3}g_{ij}\tensor{T}{^l_l}\ .
\]

\begin{figure}[ht]
\begin{center}
\psfrag{ylabel}[cc][cl][2]{\parbox{2cm}{\begin{center}$\displaystyle\eta^{(0)}$\\$\displaystyle(\mathrm{GeV}^3)$\end{center}}}
\psfrag{xlabel}[][][2]{$T\ (\mathrm{MeV})$}
\psfrag{p4}[l][l][2]{$\mathcal{O}(p^4)$}
\psfrag{unit}[l][l][2]{unitarized $\mathcal{O}(p^4)$}
\psfrag{Davesne}[l][l][2]{[Davesne] $1^\mathrm{st}$ order}
\psfrag{felz001}[l][l][2]{[DLE] $z=0.01$}
\psfrag{felz05}[l][l][2]{[DLE] $z=0.5$}
\resizebox{0.38\textwidth}{!}{
  \includegraphics{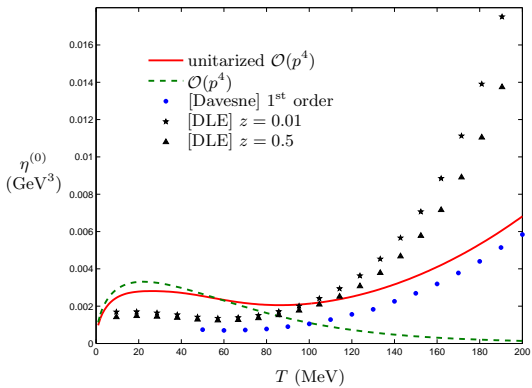}
}
\end{center}
\vspace{-0.3cm}\caption{Behavior of the unitarized shear viscosity
with temperature and comparison with the results obtained in
\cite{Viscosity,Davesne}. }\label{shearplot}
\end{figure}

The shear viscosity has been previously  calculated for a meson gas
in the KT framework \cite{Viscosity,Davesne,pra93}. For instance, in
the work \cite{Viscosity} (referred as DLE), a treatment based on a
relativistic version of the Boltzmann equation is followed, using
unitarized amplitudes from ChPT. In Fig. \ref{shearplot} we show our
leading order result for $\eta$, where we see that unitarity also
makes the shear viscosity  change its behavior with $T$. Following
the same steps as in \cite{Conductivity}, we readily get the
behavior for $T\ll M_\pi$   as $\eta\simeq 37 F_\pi^4
\sqrt{T}/M_\pi^{3/2}$ ($F_\pi\simeq 93\ \mathrm{MeV}$ is the pion decay
constant). This behavior  is consistent with nonrelativistic KT
\cite{Viscosity}.

In Fig. \ref{shearplot} we compare our leading order result with
\cite{Viscosity,Davesne}. The analysis in \cite{Viscosity} is
performed for fixed fugacity $\displaystyle
z\equiv\mathrm{e}^{\beta(\mu_\pi-M_\pi)}$. We work at $\mu_\pi=0$,
so that for the two values of $z$ shown in Fig. \ref{shearplot}, we
must compare at temperatures $T\simeq 30\ \mathrm{MeV}$ (for
$z=0.01$) and $T\simeq 200\ \mathrm{MeV}$ (for $z=0.5$). We get a
reasonable agreement with \cite{Davesne} and a lower $\eta$ than
\cite{Viscosity} for $T>M_\pi$. The results  in \cite{pra93}, where
kaons are also included in the gas, also agree with ours for low and
moderate temperatures. Nevertheless, our results should be taken
with care, since for the shear viscosity, ladder diagrams with
derivative vertices could become important already at $T\simeq 50\
\mathrm{MeV}$.

\begin{figure}[ht]
\begin{center}
\psfrag{ylabel}[cc][cl][2]{$\displaystyle\frac{\eta^{(0)}}{s}$}
\psfrag{xlabel}[][][2]{$T\ (\mathrm{MeV})$}
\psfrag{unit}[l][l][2]{unitarized $\mathcal{O}(p^4)$}
\psfrag{bound}[l][l][2]{$1/(4\pi)$} \resizebox{0.38\textwidth}{!}{
  \includegraphics{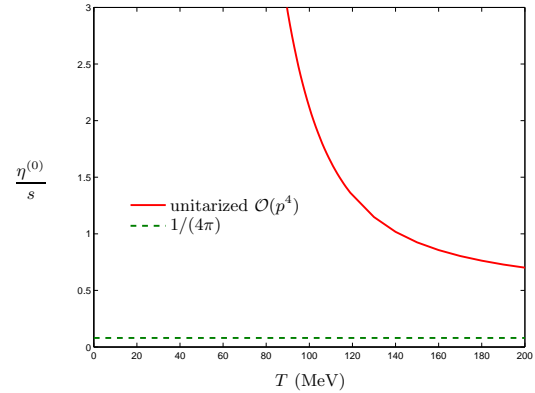}
}
\end{center}
\vspace{-0.3cm}\caption{Dependence of the $\eta/s$ ratio with
temperature in ChPT. The entropy density $s$ is calculated to
$\mathcal{O}(p^5)$ \cite{Gerber}. }\label{etasplot}
\end{figure}

An interesting quantity is the ratio $\eta/s$, where $s$ is the
entropy density. It has been conjectured \cite{Kovtun} that the
value $1/(4\pi)$ is a universal lower bound for this quantity in
any physical system. On the other hand, the sound attenuation length
$\Gamma_s\equiv 4\eta/(3s T)$ (neglecting bulk viscosity) enters
directly in phenomenological effects observed in Relativistic Heavy
Ion Collisions, such as elliptic flow or HBT radii \cite{Teaney}. In
Fig. \ref{etasplot} we plot this ratio for the pion gas. We  see that
in our analysis $\eta/s$ decreases monotonously,  but it respects
the KSS bound for temperatures up to the chiral phase transition
$T_c\simeq 180-200\ \mathrm{MeV}$. On general grounds, one expects $\eta/s$ to
decrease below $T_c$ and increase logarithmically for very high
temperatures \cite{kap}. The numerical values we get for
$\Gamma_s$($\simeq 1.1\ \mathrm{fm}$ at $T=180\ \mathrm{MeV}$) are  in remarkable
agreement with those used in \cite{Teaney} and at high $T$ our curve
is not far from recent lattice and model estimates giving
$\eta/s\sim 0.4-0.5$ above $T_c$ \cite{nasa05}.

\vspace{0.2cm}In conclusion, we have presented a systematic method
for evaluating transport coefficients in ChPT. The standard ChPT has
to be modified, and as we go to higher temperatures, more diagrams
have to be taken into account and eventually resummed. Even though
the limitations of our approach are important, we get a reasonable
agreement for physical quantities such as the very low $p_T$ photon
spectrum or the viscosity to entropy ratio. We are currently
analyzing the high-temperature diagrams, as well as the inclusion of
kaons and other transport coefficients such as bulk viscosity and
thermal conductivity. Detailed results will be reported elsewhere.

\vspace{-0.2cm}\begin{acknowledgement} We are grateful to A. Dobado
and F. Llanes-Estrada for their useful comments. We also acknowledge
financial support from the Spanish research projects FPA2004-02602,
 PR27/05-13955-BSCH, FPA2005-02327 and  FPI fellowship BES-2005-6726.
\end{acknowledgement}

\end{fmffile}

\begin{thebibliography}{}
\bibitem{Gasser}
J. Gasser and H. Leutwyler, Ann. Phys. \textbf{158}, (1984) 142.
\bibitem{LeBellac}
M. Le Bellac, \textit{Thermal field theory} (Cambridge Univ. Press, 2000).
\bibitem{Jeon}
S. Jeon, Phys. Rev. D \textbf{52}, (1995) 3591-3642.
\bibitem{Valle}
M. A. Valle Basagoiti, Phys. Rev. D \textbf{66}, (2002) 045005.
\bibitem{Goity}
J. L. Goity and H. Leutwyler, Phys. Lett. B \textbf{228}, (2002) 517.
\bibitem{Conductivity}
D. Fern\'andez-Fraile and A. G\'omez Nicola, Phys. Rev. D \textbf{73}, (2006) 045025.
\bibitem{iam}
  A. G\'omez Nicola and J. R. Pelaez, Phys. Rev. D \textbf{65}, (2002) 054009.
\bibitem{thermalrho}
 A. Dobado, A. G\'omez Nicola, F. J. Llanes-Estrada and J. R. Pelaez, Phys. Rev. C \textbf{66}, (2002) 055201.
\bibitem{Arnold}
P. Arnold, G. D. Moore and L. G. Yaffe, JHEP \textbf{0011}, (2000) 001.
\bibitem{WA98}
M. M. Aggarwal \textit{et al.} (WA98 Collaboration), Phys. Rev. Lett. \textbf{93}, (2004) 022301.
\bibitem{Rapp}
S. Turbide, R. Rapp and C. Gale, Phys. Rev. C \textbf{69}, (2004) 014903;
W. Liu and R. Rapp, \texttt{nucl-th/0604031}.
\bibitem{Viscosity}
A. Dobado and F. J. Llanes-Estrada, Phys. Rev. D \textbf{69}, (2004) 116004.
\bibitem{Davesne}
D. Davesne, Phys. Rev. C \textbf{53}, (1996) 3069-3084.
\bibitem{pra93} M. Prakash, M. Prakash, R. Venugopalan and G. M. Welke,
Phys. Rev. Lett. \textbf{70}, (1993) 1228.
\bibitem{Kovtun} P. Kovtun,
D. T. Son and A. O. Starinets, Phys. Rev. Lett. \textbf{94}, (2005)
1 11601.
\bibitem{Teaney} D. Teaney, Phys. Rev. D \textbf{68}, (2003) 034913.
\bibitem{kap} L. P. Csernai, J. I. Kapusta and L. D. McLerran, Phys. Rev. Lett. \textbf{97}, (2006) 152303.
\bibitem{nasa05} A. Nakamura and S. Sakai, Phys. Rev. Lett. \textbf{94},
(2005) 72305; B. A. Gelman, E. V. Shuryak and I. Zahed,
\texttt{nucl-th/0601029}.
\bibitem{Gerber} P. Gerber and H. Leutwyler, Nucl. Phys. B
\textbf{321}, (1989) 387.
\end{thebibliography}
\end{document}